# From virtual work principle to maximum entropy for nonequilibrium system

Qiuping A. Wang

Institut Supérieur des Matériaux et Mécaniques Avancés du Mans, 44 Av. Bartholdi,

72000 Le Mans, France

Abstract

After the justification of the maximum entropy approach for equilibrium thermodynamic system, and of a maximum path entropy algorithm for nonequilibrium thermodynamic systems by virtue of the principle of virtual work, we present in this paper another application of the principle to thermodynamic systems out of equilibrium. Unlike the justification of maximum path entropy for the motion trajectories during a period of time, this work is on the maximum of the entropy defined as a measure of the momentary dynamical uncertainty as a function of the probability distribution over the microstates of the system at any given moment.

## 1) Introduction

The maximum entropy principle (maxent) proposed by Jaynes[1] for equilibrium system represents a longstanding belief dating back to Boltzmann and Gibbs. As presented in many textbooks, one can use the Boltzmann entropy in logarithmic function of the number of microstates of an isolated system as well as equal probability for all the microstates, then the assertion that the maximum entropy state is the most realist one looks reasonable since the most probable state must contain in this case of equipartiton the largest number of microstates which imply largest entropy. However, the extension of this assertion to other ensembles (canonical ensemble for example), and especially the arguments for this extension, are less obvious with the expression of thermodynamic entropy in Shannon formula ($S=-\sum_i p_i \ln p_i$) where the probability $p_i$ are different in general for different microstates. This formula is claimed by Jaynes to be the only maximizable information measure for any probabilistic situation. In any case, as a principle, maxent does not has sound base with, for example, mathematical justification. So it seems not a physical principle on its own. Without the a priori assumption of entropy form, maxent is only supported by arguments which are either philosophical or intuitive and often formulated on the basis of the subjective nature of probability[1], in contrast to the objective nature of the frequency definition of probability in physics.

The extension of maxent to nonequilibrium system is much less obvious. This belief stems directly from the success of the maxent within equilibrium statistical mechanics and from the arguments supporting maxent as a generic inference principle for probability assignment. Since the thermodynamic entropy of the second law does not exist any more, there are different definition of entropy in this practice. The first is an extension of the equilibrium entropy to nonequilibrium system as a function of nonequilibrium state (see for example [2]). The second concerns not the entropy itself but the entropy production used in a similar maximum principle[3]. The notion of entropy production is different in different context of entropy. It is sometimes related to heat production of the process, sometimes not[4]. The third is the path entropy defined as a measure of the uncertainty in the choice (by the system) of different possible trajectories from one state to another. So it is a function of the probabilities of the occurrence of different paths. The maximization of this entropy and its consequence for the maximization of the first entropy or the entropy production has been recently studied independently by several authors[5][6][7].

In spite of the successful application of maxent to many kinds of system and of the strong and longstanding belief in its validity especially for equilibrium system, controversy and doubt persist around it (see e.g., [1][3][8]). The doubts run especially around the intuitive, plausible and non mathematical arguments supporting maxent as a generic principle for optimization, inference and deduction using a unique Shannon entropy measure.

In our opinion, a justification of maxent from other fundamental principle of physics, especially from obvious and widely accepted principles, might help to understand and to interpret maxent from theoretical and philosophical viewpoint. This is the aim of our two recent works: one for maxent of equilibrium system using the thermodynamic entropy[9], another for maxent for nonequilibrium system using path entropy[10]. Both works are based on a single principle of the classical mechanics: the principle of virtual work. The latter is an obvious, palpable, widely accepted and successfully used principle in both analytical mechanics theory and mechanical engineering for static equilibrium and dynamical equilibrium problems. For the time being, the outcome of these works is the following. 1) Maximum thermodynamic entropy is kind of dynamical equilibrium prescribed by the virtual work principle. 2) This maxent does not need any assumption about the entropy property and functional form. 3) The equiprobability situation for microcanonical ensemble is a natural consequence of this approach whatever the entropy form. 4) The constraints introduced as partial knowledge into maxent by Jaynes on the basis of informational arguments appear here naturally as a consequence of vanishing virtual work. For example, for canonical ensemble, virtual work is equal to the virtual variation of the difference between the entropy ($S$) and the energy ($\overline{E}$) at inverse temperature $\beta$, meaning that $(S - \beta\overline{E})$ should be maximized. For grand-canonical ensemble, the maximizable quantity is $(S - \beta\overline{E} + \beta\mu N)$ where $\mu$ is the potential energy and $N$ the particle number. 5) For nonequilibrium system, the maximum quantity is the difference $(S_{ab} - \eta \overline{A}_{ab})$ between the path entropy $S_{ab}$ and the mean Lagrange action $\overline{A}_{ab}$ of mechanics averaged over the different trajectories, where the path entropy is defined by $dS_{ab} = \chi(d\overline{A}_{ab} - \overline{dA_{ab}})$ ($\chi$ is a diffusion constant) without other conditions.

In the present work, we will present a further development of this approach: its application to nonequilibrium system in order to justify maxent with entropy defined as a measure of the dynamical uncertainty at any moment instead of the path entropy measuring the uncertainty of a period of the motion.

## 2) Principle of virtual work

In mechanics, a virtual displacement of a system is a kind of hypothetical infinitesimal displacement with no time passage and no influence on the forces. It should be perpendicular to the constraint forces. The principle of virtual work says that the total work done by all forces acting on a system in static equilibrium is zero for any possible virtual displacement. Suppose a simple case of a system of $N$ points of mass in equilibrium under the action of $N$ forces $F_i$ ($i=1,2,…N$) with $F_i$ on the point $i$, and imagine virtual displacement of each point $\delta \vec{r}_i$ for the point $i$. According to the principle, the virtual work $\delta W$ of all the forces $F_i$ on all $\delta \vec{r}_i$ vanishes for static equilibrium, i.e. $\delta W = \sum_{i=1}^{N} \vec{F}_i \cdot \delta \vec{r}_i = 0$ [11]. This principle was extended to moving system by d'Alembert[12] who added the initial force $-m_i \vec{a}_i$ on each point of the system where $m_i$ is the mass of the point $i$ and $\vec{a}_i$ its acceleration. This "dynamical equilibrium" is given by $\delta W = \sum_{i=1}^{N} (\vec{F}_i - m_i \vec{a}_i) \cdot \delta \vec{r}_i = 0$.

## 3) Maximum entropy for nonequilibrium system

We have an ensemble of a large number of identical systems out of equilibrium. For any moment of the movement, the systems are distributed over the momentary time dependent microstates in the same way as an ensemble of equilibrium systems distributed over the time independent microstates. Suppose in this ensemble a system is composed of $N$ particles moving in the $3N$ dimensional position space starting from a point $a$. If the motion was regular, all the systems would follow a single trajectory from $a$ to a given point $b$ in the $6N$-dimensional phase space during a given period $\tau$ according to the least action principle. But as well known every system is subject to irregular motion due to the random motion of the particles. This random dynamics is just the underlying reason for the fluctuation in energy and other quantity of the system. In this circumstance, the systems may take different paths in the phase space from $a$ to $b$ as discussed in the papers [5][6][7]. Without loss of generality, we consider discrete paths denoted by $k=1,2$ … (if the variation of the paths is continuous, the sum over $k$ must be replaced by path integral between $a$ and $b$). At a given moment $t$ after the departure of the ensemble from an initial state, all the systems are distributed over the different trajectories, each one arriving at a microstate $j$ ($j=1,2,$ … w) . This implies that at that moment, each trajectory arrives at a microstate. Let $p_j$ be the probability that a system is found at the state *j* having a quantity $x_j$. The ensemble average of the quantity is then given by

$$\bar{x} = \sum_{j=1}^{w} x_j p_j .$$

Now let us look at the random dynamics of a single system. After leaving its initial state, the system is found at a moment on the trajectory *k* or at the corresponding microstate *j*. At this time, the total force on a particle *i* of the system is denoted by $\vec{F}_i$ and the acceleration by $\vec{a}_i$ with an inertial force $-m_i \vec{a}_i$ where $m_i$ is the mass of the particle. Suppose a virtual displacement $\delta \vec{r}_i$ of the particle *i* at the state *j*, the virtual work over this displacement is given by

$$\delta W_{ij} = (\vec{F}_i - m_i \vec{a}_i)_j \cdot \delta \vec{r}_i \qquad (1)$$

Summing this work over all the particles, we obtain

$$\delta W_j = \sum_{i=1}^{N} (\vec{F}_i - m_i \vec{a}_i)_j \cdot \delta \vec{r}_i \qquad (2)$$

For the second terms at the right hand side with inertial forces, we can calculate $m\vec{a}_i \cdot \delta \vec{r}_i = m\delta \dot{\vec{r}}_i \cdot \dot{\vec{r}}_i = \delta(\frac{1}{2} m\dot{\vec{r}}_i^2) = \delta t_i$ where $t_i$ is the kinetic energy of the particle *i*. To treat the first terms with $\vec{F}_i$, we will distinguish the forces of different nature. Let $\vec{f}_i$ be the interaction forces between the particles whose total vanish for the global system just as in the case of equilibrium system. These forces can be given by $\vec{f}_i = -\nabla w_i$ where $w_i$ is the one particle potential energy. Let $\varphi_{il}$ be the forces due to the gradient of some other variables such as pressure, temperature, particle density, external fields, etc (with *l*=1,2, …). These forces can be given by $\vec{\varphi}_{il} = -\mu_l \nabla_i V_l$ where $V_l$ is the effective potential and $\mu_l$ is some constant uniquely related to the nature of each force $\vec{\varphi}_{il}$ or $V_l$. We can write

$$\sum_{i=1}^{N} \vec{F}_i \cdot \delta \vec{r}_i = -\sum_i (\nabla w_i + \sum_l \mu_l \nabla_i V_l) \cdot \delta \vec{r}_i = -\sum_{i=1}^{N} (\delta w_i + \sum_l \mu_l \delta v_{il}) . \qquad (3)$$

where $\delta v_{il} = \nabla_i V_l \cdot \delta \vec{r}_i$ is the virtual variation in the potential $V_l$ due to the virtual displacement of the particle *i*. It follows that

$$\delta W_j = -\sum_{i=1}^{N} (\delta w_i + \delta t_i)_j - \sum_{i=1}^{N} \sum_l \mu_l \delta v_{il} = -\sum_{i=1}^{N} (\delta e_i + \sum_l \mu_l \delta v_{il})_j . \qquad (4)$$

where $\delta e_i$ is the virtual variation of only a part ($e_i = t_i + w_i$) of the total energy of the particle for the microstate $j$. This is in fact the one particle energy when the system is in equilibrium since $w_i$ is defined as the potential of the equilibrium forces. In this way, the virtual work can be separated into two part: $\delta W_j = \delta W_j^{eq} + \delta W_j^{neq}$, the first is the virtual work for equilibrium state $\delta W_j^{eq} = -\sum_{i=1}^{N}(\delta e_i)_j$ and the second is for nonequilibrium states $\delta W_j^{neq} = -\sum_{i=1}^{N}(\sum_l \mu_l \delta v_{il})_j$.

Remember that, from microscopic viewpoint, a microstate $j$ is a distribution $\{n_1, .. n_k, … n_g, \}_j$ of the $N_j$ particles of a system over the $g$ one-particle states $k=1,2 …g$ each having energy $e_k$. So we can change the sum over $i$ into the sum over $k$. For equilibrium work, one has

$$\delta W_j^{eq} = -\delta \sum_{k=1}^{g}(n_k e_k)_j = -\sum_{k=1}^{g}(n_k \delta e_k - e_k \delta n_k)_j. \tag{5}$$

Hence the equilibrium part of the total virtual work is

$$\delta W^{eq} = \sum_j p_j \delta W_j^{eq} = -\sum_j p_j \sum_k (n_k \delta e_k)_j - \sum_j p_j \sum_k (e_k \delta n_k)_j \tag{6}$$
$$= -\sum_k \delta e_k \sum_j p_j n_{kj} - \sum_k e_k \sum_j p_j \delta n_{kj} = -\sum_k \overline{n_k} \delta e_k - \sum_k e_k \overline{\delta n_k}.$$

The first term is the average of the virtual variation of the energy of the system with constant particle number, which can be denoted by $\overline{\delta E} = \sum_k \overline{n_k} \delta e_k = \sum_j p_j \delta E_j$ with $\delta E_j = \sum_k (n_k \delta e_k)_j$. The second term is the virtual variation of the system energy due to changing particle number (grand-canonical system), which is $-\sum_k e_k \overline{\delta n_k} = \mu \sum_j p_j \delta N_j = \mu \overline{\delta N}$ with $\mu \delta N_j = \sum_k (e_k \delta n_k)_j$ and $\mu = -\sum_k e_k \overline{\delta n_k} / \overline{\delta N}$ (chemical potential). Since $\overline{\delta E} = \sum_{j=1}^{w} p_j \delta E_j = \delta \sum_{j=1}^{w} p_j E_j - \sum_{j=1}^{w} E_j \delta p_j$ and $\overline{\delta N} = \delta \sum_{j=1}^{w} p_j N_j - \sum_{j=1}^{w} N_j \delta p_j$, we finally get

$$\delta W^{eq} = -\delta \overline{E} + \mu \delta \overline{N} + \sum_j (E_j - \mu N_j) \delta p_j. \tag{7}$$

By virtue of the first law of thermodynamics, we can identify the heat transfer $\delta Q = \sum_j (E_j - \mu N_j) \delta p_j$. If we suppose a reversible virtual displacement (not to be confused with the irreversible nonequilibrium process under consideration), we can still write $\delta Q = \delta S^{eq}/\beta$.

Now let us see the nonequilibrium part of the total virtual work $\delta W_j^{neq} = -\sum_{i=1}^{N}(\sum_l \mu_l \delta v_{il})_j$. Its expression over the one particle states is

$$\delta W_j^{neq} = -\delta \sum_{k=1}^{g}(n_k \sum_l \mu_l v_{kl})_j = -\sum_{k=1}^{g}(n_k)_j \sum_l \mu_l \delta v_{kl} - \sum_{k=1}^{g}\sum_l \mu_l v_{kl}(\delta n_k)_j. \tag{8}$$

Then the average nonequilibrium virtual work is given by

$$\delta W^{neq} = \sum_{j=1}^{g} p_j \delta W_j^{neq} = -\sum_l \mu_l (\sum_{k=1}^{g} \overline{n_k} \delta v_{kl} - \sum_{k=1}^{g} v_{kl} \overline{\delta n_k}). \tag{9}$$

For a given $l$, the first term in the parentheses is in fact $-\sum_j p_j \sum_{k=1}^{g}(n_k \delta v_{kl})_j = -\sum_j p_j \delta V_{lj} = -\overline{\delta V}_l$ where $\delta V_{lj} = \sum_k n_k \delta v_{kl}$ is the sum of the virtual variations of $v_{kl}$ over all the particles (or over all the one particle states) at the state $j$. The second term can be written as $-\sum_j p_j \sum_{k=1}^{g} v_{kl}(\delta n_k)_j = -v_l \sum_j p_j \delta N_j = -v_l \overline{\delta N}$ where $v_l = \sum_{k=1}^{g} v_{kl}(\delta n_k)_j / \delta N_j$. It follows that

$$\delta W^{neq} = \sum_{j=1}^{g} p_j \delta W_j^{neq} = -\sum_l \mu_l (\overline{\delta V}_l + v_l \overline{\delta N}) = -\sum_l \mu_l \overline{\delta V}_l - \omega \overline{\delta N}. \tag{10}$$

where $\omega = \sum_l \mu_l v_l$ can be considered as a kind of nonequilibrium chemical potential. Since $\overline{\delta V}_l = \delta \sum_{j=1}^{w} p_j V_{lj} - \sum_{j=1}^{w} V_{lj} \delta p_j$, Eq.(10) becomes

$$\delta W^{neq} = -\sum_l \mu_l \delta \overline{V}_l + \omega \delta \overline{N} + \sum_j (\sum_l \mu_l V_{lj} - \omega N_j) \delta p_j. \tag{11}$$

Mimicking the first law, we can define the last term of this equation as a kind of "heat" implying that it is associated with the uncertainty of the probability distribution as a function of $V_{lj}$ and $N_j$ in the nonequilibrium state. Let it be denoted by

$$\delta S^{neq} = \eta \sum_j (\sum_l \mu_l V_{lj} - \omega N_j) \delta p_j \tag{12}$$

where $\eta$ is a nonequilibrium parameter which mimics $\beta$ for equilibrium state. Finally, the total virtual work for the nonequilibrium system is

$$\delta W = -\delta \overline{E} - \sum_l \mu_l \delta \overline{V}_l + (\mu+\omega)\delta \overline{N} + \delta S^{neq}/\eta + \delta S^{eq}/\beta \qquad (13)$$

$$= -\delta \overline{E} - \sum_l \mu_l \delta \overline{V}_l + (\mu+\omega)\delta \overline{N} + \delta \Omega$$

where $\delta \Omega = \delta S^{neq}/\eta + \delta S^{eq}/\beta$ is the total uncertainty of the dynamics. Applying the virtual work principle to this equation gives the following variational approach

$$\delta \Omega - \delta \overline{E} - \sum_l \mu_l \delta \overline{V}_l + (\mu+\omega)\delta \overline{N} = 0 \qquad (14)$$

for nonequilibrium dynamics. In what follows, we only discuss the case of closed system without variation of particle number, i.e., $\mu=\nu_l=\omega=0$ for canonical ensemble. Eq.(14) is now:

$$\delta(S^{eq}/\beta + S^{neq}/\eta - \overline{E} - \sum_l \mu_l \overline{V}_l) = 0, \qquad (15)$$

a global maxent for the nonequilibrium dynamics.

## 4) Maximum entropy production

We remember that the dynamics is separated into two components: equilibrium one and nonequilibrium one. This separation makes it possible to tackle the dynamical uncertainty and total energy in two components as well: the equilibrium uncertainty associated with the second law $\delta S^{eq} = \beta \delta Q$ and the nonequilibrium uncertainty $\delta S^{neq} = \eta \delta Q^{neq}$; the equilibrium internal energy $\overline{E}$ and the nonequilibrium internal energy $\sum_l \mu_l \overline{V}_l$. This separation can help us to understand more about the variational recipe of Eq.(14). In a previous work[9], we have proved that for equilibrium system, the variation of the difference $(S^{eq} - \beta \overline{E})$ vanishes, which implies a variational algorithm for the nonequilibrium component only

$$\delta(S^{neq} - \eta \sum_l \mu_l \overline{V}_l) = 0. \qquad (16)$$

This means that the nonequilibrium virtual work must vanish as well $\delta W^{neq} = 0$. So it is $(S^{neq} - \eta \sum_l \mu_l \overline{V}_l)$ which is maximized at any moment of the dynamics. From point of view of the Jaynes' maxent algorithm, this is just the maximum of the nonequilibrium entropy under the constraint of the effective potentials $\mu_l \overline{V}_l$.

In order to see further the maxent of $S^{neq}$, let us suppose the nonequilibrium entropy is a function of the probability distribution $p_j$ at the considered moment, i.e., $S^{neq} = f(p_1, p_2, ... p_w)$. We can write $\delta S^{neq} = \sum_j \frac{\partial f}{\partial p_j} \delta p_j$ due to a virtual process. On the other hand, we have $\delta S^{neq} = \eta \sum_j \sum_l \mu_l V_{lj} \delta p_j$ which implies

$$\sum_j (\frac{\partial f}{\partial p_j} - \eta \sum_l \mu_l V_{lj}) \delta p_j = 0. \tag{17}$$

By virtue of the normalization condition $\sum_{j=1}^{w} \delta p_j = 0$ for this given moment, one can prove that

$$\frac{\partial f}{\partial p_j} - \eta \sum_l \mu_l V_{lj} = K. \tag{18}$$

with constant $K$[14]. Eq.(18) can be used for deriving the probability distribution of the nonequilibrium component of the dynamics if the functional $f$ is given. Inversely, if the probability distribution is known, one can derive the functional of $S^{neq}$ (some examples are given in [13]).

Concerning Eq.(14), since $\mu_l \overline{V}_l$ is the potentials associated with macroscopic transfer such as matter diffusion, heat conduction, macroscopic expansion of gas etc, $\delta S^{neq}$ can be seen as the entropy (momentary uncertainty) of the transfer processes. Without this transport phenomenon, it must be an equilibrium, $S^{neq}$ must be constant. From this viewpoint, $S^{neq}$ is an entropy production of the transports, and Eq.(14) is in fact the maximization of the entropy production (MEP) or of the uncertainty produced by the dynamics. MEP is a principle proposed since long for nonequilibrium thermodynamics and verified by numerous systems (see [3] and references there-in for details).

One should notice the difference between the entropy in MEP and the thermodynamic entropy used for equilibrium case. We do not have necessarily $\delta S^{neq} = \beta \delta Q$ where $\delta Q$ is the heat transfer.

5) Application to diffusion

In what follows, we do not consider energy dissipation. We address simple cases with only one effective potential $V_l$ for transfer phenomena.

Let $C$ be the density of the diffused particles. From the Fick's first law, the flux is given by $J = -D\nabla C$ where $D$ is the diffusion constant. Comparing the law to $J = Cu = C\mu_d F$ ($u = \mu_d F$ is the drift velocity of a particle with $\mu_d$ its mobility and $F$ the force on each particle), we obtain the effective force $\bar{\varphi}_{il} = F = \dfrac{-D\nabla C}{C\mu_d}$. We can see that in the definition $\bar{\varphi}_{il} = -\mu_l \nabla_i V_l$, the constant $\mu_l = \dfrac{D}{\mu_d}$ if we let $\nabla = \nabla_i$ and $V_l = \bar{V}_l = \ln C$. We get the entropy production

$$\delta S^{neq} = \eta \mu_l \delta \ln C. \tag{19}$$

This result can be applied to the free adiabatic expansion of ideal gas of $N$ particles. Suppose a small expansion from a state 1 to a state 2 with respectively volumes $V_1$ to $V_2$ and densities $C_1$ and $C_2$. It is straightforward to write $\delta S^{neq} = \eta \mu_l \ln \dfrac{V_2}{V_1}$. This is close to the result $\Delta S = N k_B \ln \dfrac{V_2}{V_1}$ calculated for two equilibrium states 1 and 2. If it is the case, we can identify $\eta \mu_l$ to $Nk_B$ where $k_B$ is the Boltzmann constant, which means the following relationship $\eta = \dfrac{\mu_d N k_B}{D}$. Using the Einstein relation $D = \mu_d k_B T$, we find $\eta = \dfrac{N}{T}$. Note that $\Delta S$ is the difference between the two equilibrium entropies of the state 1 and 2. But Eq.(19) is in general for any two momentary states during the expansion.

In the case of electric conduction with the current density $\vec{J} = -Ce\vec{u} = -Ce\mu_d \vec{E}$ with $\vec{E} = -\nabla V$, where $e$ is the charge, $\vec{E}$ the electric field and $V$ its potential, we have $\bar{\varphi}_{il} = -e\nabla V$ with $\mu_l = e$ and $V_l = V$. Hence

$$\delta S^{neq} = \eta e \delta V. \tag{20}$$

By virtue of the Joule's law $P = \dfrac{\delta Q}{\delta t} = I\delta V$ with current intensity $I$ and heat production $\delta Q$ over a small segment $\delta l$ of the conduction wire crossed by the current during a time period $\delta t$, we can identify $T\delta S^{neq}$ to $\delta Q$ if the temperature is uniform everywhere in the volume $s\delta l$ of the segment $\delta l$ where $s$ is the section area of the wire. Since $\delta Q = P\delta t = Ceus\delta V\delta t = Ne\delta V$ where $N = Csu\delta t = Cs\delta l$ is the total number of electrons in the segment $\delta l$. Put $\delta Q = T \delta S^{neq}$ and compare it with Eq.(20), we find $\eta = \frac{N}{T}$ as above and Eq.(20) becomes $\delta S^{neq} = \frac{Ne}{T} \delta V$.

We still notice that even within this analysis based on the heat production, $\delta S^{neq}$ is not the variation of the thermodynamic entropy of second law which does not exist for nonequilibrium system.

6) Concluding remarks

On the basis of the application of virtual work principle of d'Alembert to equilibrium system in order to justify maximum thermodynamic entropy, we have presented here an extension of the same principle to thermodynamic system out of equilibrium in order to justify maxent using the momentary entropy as a measure of the dynamical uncertainty at any moment instead of the path entropy as the uncertainty of trajectory for a period of the evolution. We see that maxent for nonequilibrium system with the momentary entropy as a function of the probability distribution over the momentary microstates is a consequence of a very basic principle in mechanics for equilibrium problems.

One outcome of this work is that maxent, being mathematically justified from other physical principle, is a physics principle on its own.

Another conclusion is that the validity of this maxent does not need hypothesis about the entropy functional. This opens the way for maxent to be applied to other entropy forms if any as measure of statistical uncertainty. See reference [13] for detailed discussion on this topic.